\documentclass[acus]{JAC2000}

\usepackage{graphicx}

\begin{document}
\title{\flushright{W08}\\[10pt] 
       \centering NUCLEON FORM FACTORS IN THE SPACE- AND TIMELIKE  REGIONS}

\author{H.-W. Hammer, The Ohio State University, Columbus, OH 43210, USA}

\maketitle

\begin{abstract}
Dispersion relations provide a powerful tool to describe the
electromagnetic form factors of the nucleon both in the spacelike
and timelike regions with constraints from
unitarity and perturbative QCD. We give a brief introduction into 
dispersion theory for nucleon form factors and present results
from a recent form factor analysis. Particular emphasis is given 
to the form factors in the timelike region.
Furthermore, some recent results for the spacelike
form factors at low momentum transfer from a ChPT calculation
by Kubis and Mei{\ss}ner are discussed.
\end{abstract}

\section{INTRODUCTION}

A detailed understanding of the electromagnetic form factors of the
nucleon is important for revealing aspects of both perturbative and 
nonperturbative nucleon structure. The form factors also contain
important information on nucleon radii and vector meson coupling
constants. Moreover, the form factors are an
important ingredient in a wide range of experiments such as
Lamb shift measurements \cite{Ude97} and measurements of the
strangeness content of the nucleon \cite{strange}.
With the advent of the new continuous beam electron accelerators
like CEBAF (Jefferson Lab.), MIT-Bates, and MAMI (Mainz), a
wealth of precise data for spacelike momentum transfers
has become available. Due to the difficulty of the experiments, the
timelike form factors are less well known. While there is a fair 
amount of information on the proton timelike form factors,
only one measurement of the neutron form factor from the pioneering 
FENICE experiment \cite{Ant98} exists.

The basic theoretical framework of dispersion theory for nucleon
form factors was established long ago \cite{disp,Hoe76}.
In recent years, these methods have been refined considerably. 
In particular, constraints from perturbative QCD and independent 
low-energy experiments (e.g. on the neutron 
radius) have been incorporated \cite{MMD96,HMD96}.

In the first part of this paper, we review the status of
dispersion theory for nucleon form factors and show results
from a recent update \cite{MMD96,HMD96}.
In the second part, we discuss the effective field theory 
description of nucleon form factors in the framework of ChPT, based on 
a recent calculation by Kubis and Mei{\ss}ner\cite{KuM01}.

\section{FORMALISM}

Using Lorentz and gauge invariance, the nucleon matrix element of the 
electromagnetic (em) current operator $j_\mu^{\rm em}$ can be parametrized
in terms of two form factors,
\begin{equation}
\langle p' | j_\mu^{\rm em} | p \rangle = \bar{u}(p')
\left[ F_1 (t) \gamma_\mu +i\frac{F_2 (t)}{2 M} \sigma_{\mu\nu}
q^\nu \right] u(p)\,,
\end{equation}
where $M$ is the nucleon mass and $t=(p'-p)^2=q^2 <0$ 
the four-momentum transfer. 
$F_1$ and $F_2$ are the Dirac and Pauli form factors, respectively.
They are normalized at $t=0$ as
\begin{equation}
\label{norm}
F_1^p(0) = 1\,, \; F_1^n(0) = 0\,, \; F_2^p(0) =  \kappa_p\,,
\; F_2^n(0) = \kappa_n\, ,
\end{equation}
with $\kappa_p=1.79$ and $\kappa_n=-1.91$ the anomalous magnetic moments of
protons and neutrons in nuclear magnetons, respectively.
It is convenient to work in the isospin basis and to 
decompose the form factors into isoscalar and isovector parts,
\begin{equation}
F_i^s = \frac{1}{2} (F_i^p + F_i^n) \, , \quad
F_i^v = \frac{1}{2} (F_i^p - F_i^n) \, ,
\end{equation}
where $i = 1,2 \,$. 
The experimental data are usually given for the Sachs form factors
\begin{eqnarray}
\label{sachs}
G_{E}(t) &=& F_1(t) - \tau F_2(t) \, , \\
G_{M}(t) &=& F_1(t) + F_2(t) \, , \nonumber
\end{eqnarray}
where $\tau = -t/(4 M^2)$.
In the Breit frame, $G_{E}$ and $G_{M}$ may be interpreted as
the Fourier transforms of the charge and magnetization distributions,
respectively.              

\section{NUCLEON FORM FACTORS IN DISPERSION THEORY}

\subsection{Spectral Decomposition}

Based on unitarity and analyticity, dispersion relations relate
the real and imaginary parts of the em nucleon form factors. 
Let $F(t)$ be a generic symbol for any one of the four nucleon form
factors. We write down an unsubtracted
dispersion relation of the form
\begin{equation}
\label{disprel}
F(t) = \frac{1}{\pi} \, \int_{t_0}^\infty \frac{{\rm Im}\, 
F(t')}{t'-t-i\epsilon}\, dt'\, ,
\label{emff:disp} 
\end{equation}
where $t_0$ is the threshold of the lowest cut 
of $F$.
Since the normalization of $F$ is known, a once subtracted dispersion 
relation could be used as well.
Eq.~(\ref{emff:disp}) relates the em structure
of the nucleon to its absorptive behavior.
The imaginary part entering Eq.~(\ref{disprel}) 
can be obtained from a spectral decomposition \cite{disp}. 
For this purpose it is convenient to consider the 
em current matrix element in the timelike region,
\begin{eqnarray}
\label{eqJ}
J_\mu &=& \langle N(p) \overline{N}(\bar{p}) | j_\mu^{\rm em}(0) | 0 \rangle \\
&=& \bar{u}(p) \left[ F_1 (t) \gamma_\mu +i\frac{F_2 (t)}{2 M} \sigma_{\mu\nu}
(p+\bar{p})^\nu \right] v(\bar{p})\,,\nonumber
\end{eqnarray}
where $p\,,\bar{p}$ are the momenta of the nucleon-antinucleon pair
created by the current $j_\mu^{\rm em}$. The four-momentum transfer
in the timelike region is $t=(p+\bar{p})^2 > 0$. 
Using the LSZ formalism, the imaginary part
of the form factors is obtained by inserting a complete set of
intermediate states as \cite{disp}
\begin{eqnarray}
\label{spectro}
{\rm Im}\,J_\mu &=& \frac{\pi}{Z}(2\pi)^{3/2}{\cal N}\,\sum_\lambda
 \langle p | \bar{J}_N (0) | \lambda \rangle \qquad\\
& &\times\langle \lambda | j_\mu^{\rm em} (0) | 0 \rangle \,v(\bar{p})
\,\delta^4(p+\bar{p}-p_\lambda)\,,\nonumber
\end{eqnarray}
where ${\cal N}$ is a nucleon spinor normalization factor, $Z$ is
the nucleon wave function renormalization, and $\bar{J}_N (x) =
J^\dagger(x) \gamma_0$ with $J_N(x)$ a nucleon source.
The states $|\lambda\rangle$ are asymptotic states of
momentum $p_\lambda$ which are stable with respect to the strong 
interaction. For the matrix elements in Eq.~(\ref{spectro}) to be
nonvanishing, they must carry the same quantum numbers as as
the current $j^{\rm em}_\mu$: $I^G(J^{PC})=0^-(1^{--})$ for
the isoscalar component and $I^G(J^{PC})=1^+(1^{--})$ for the
isovector component of $j^{\rm em}_\mu$. 
Furthermore, they have no net baryon number.
For the isoscalar part  the lowest mass states are: $3\pi$,
$5\pi$, $\ldots$; for the isovector part they
are: $2\pi$, $4\pi$, $\ldots$. Because of $G$-parity, states
with an odd number of pions only contribute to the isoscalar
part, while states with an even number contribute to the 
isovector part. Associated with each intermediate state is a
cut starting at the corresponding threshold in $t$ and running to
infinity. As a consequence,
the spectral function ${\rm Im}\, F(t)$ is different from zero along the
cut from $t_0$ to $\infty$ with $t_0 = 4 \, (9) \, m_\pi^2$ for the
isovector (isoscalar) case. 
Using Eqs.~(\ref{eqJ},\ref{spectro}), the spectral functions for
the form factors can in principle be constructed from experimental
data. In practice, this proves a formidable task and is also unstable.
However, the spectral function can be constrained
using, for example, vector meson dominance.

\subsection{Vector Meson Dominance}

Within the vector meson dominance (VMD) approach, the spectral
functions are approximated by a few vector meson poles, namely the $\rho,
\ldots$ in the isovector and the $\omega, \phi, \ldots$ in the
isoscalar channel, respectively. In that case, the form factors take the
form
\begin{equation}
F_i^{I} (t) = \sum_{V_I} \frac{a_i^{V_I}}{m^2_{V_I}-t}
\, \quad i = 1,2 \, ; \,  \quad I = v,s  \, \, . 
\label{emff:vmd}
\end{equation}
Clearly, such pole terms contribute to the spectral functions
as $\delta$-functions,
\begin{equation}
{\rm Im}\, F^{V_I}_i (t) = \pi\,  a_i^{V_I} \, 
\delta(t - m_{V_I}^2) \, \, .
\end{equation}
These terms arise naturally as approximations to 
vector meson resonances in the continuum
of intermediate states like $n \pi$ ($n \geq 2)$, 
$N \overline{N}$, $K \overline{K}$
and so on. If the continuum contributions are strongly peaked near
the vector meson resonances, Eq. (\ref{emff:vmd}) is a good
approximation. 

For the contribution of the
two-pion continuum to the isovector spectral functions, however, the
replacement by a sharp $\rho$-resonance is not justified
and strongly underestimates the isovector radius. 
The unitarity relation of Frazer and Fulco
\cite{FrF60} determines the isovector spectral functions from $t=
4\,m_\pi^2$ to $t \simeq 50\, m_\pi^2 \simeq 1$ GeV$^2$ in terms of the
pion form factor $F_\pi (t)$ and the P-wave $\pi \pi N \bar{N}$
partial wave amplitudes.
The isovector spectral function is strongly enhanced close to the 
two-pion threshold (see, e.g., Fig. 2 in Ref. \cite{MMD96}).
The reason for this behavior is well known. The
corresponding partial waves have a branch point singularity on the
second sheet (from the projection of the nucleon pole terms) located
at $t_c = 4 m_\pi^2 -m_\pi^4 / M^2  \approx 3.98 \, m_\pi^2 \,,$
very close to the physical threshold at $t_0 = 4 m_\pi^2$. The
isovector form factors inherit this singularity and the closeness to
the physical threshold leads to the pronounced enhancement. Note that
in the VMD approach this spectral function is given by a
$\delta$-function peak at $m_\rho^2 \simeq 30\, m_\pi^2$ and thus the
isovector radii are strongly underestimated if one neglects the
unitarity correction \cite{HoP75a}, as can be seen from the formula
\begin{equation}
\langle r^2\rangle_i^{v} = \frac{6}{\pi}
 \int_{4m_\pi^2}^\infty \frac{dt}{{t}^2} \, {\rm Im}\,
  F_i^{v} (t) \, \, . 
\end{equation}
In the isoscalar channel, it is believed that
the pertinent spectral functions rise smoothly
from the three-pion threshold  to the $\omega$-peak,
i.e. that there is no pronounced effect from the three-pion cut on
the left wing of the $\omega$-resonance (which also has a much
smaller width than the $\rho$).
In chiral perturbation theory an investigation of the spectral functions
to two loops has indeed shown no such enhancement \cite{BKM96},
in contrast to the one loop calculation of the isovector nucleon form
factors where the unitarity correction on the left wing of the $\rho$ has
been seen. A form factor analysis in the described framework with
three vector meson poles in both the isoscalar and isovector channels
as well as the two-pion continuum was carried out by H{\"o}hler et
al. \cite{Hoe76}. In the following, we describe an extended
update that also includes timelike data and additional contstraints 
on the spectral function.

\subsection{Constraints}

The first set of constraints concerns the low-$t$ behavior of the
form factors.
First, we enforce the correct normalization of the form factors,
which is given in Eq.~(\ref{norm}). Second, we constrain the 
neutron radius from a low-energy neutron-atom scattering experiment
\cite{Kop95}.\footnote{There has been some controversy about this value 
recently and for future analyses the error of this constraint
should be enlarged.}

Perturbative QCD (pQCD) constrains the behavior of the nucleon
em form factors for large momentum transfer.
Brodsky and Lepage \cite{BrL80} find for $t \to -\infty$,
\begin{equation}
F_i (t) \to (-t)^{-(i+1)} \, \left[ \ln\left(\frac{-t}{Q_0^2}\right)
\right]^{-\gamma} \, , \quad i = 1,2 \, ,
\label{emff:fasy1}
\end{equation}
where $Q_0 \simeq \Lambda_{\rm QCD}$.
The anomalous dimension $\gamma$ depends weakly on the number of
flavors, $\gamma = 2.148$, $2.160$, $ 2.173$ for $N_f = 3$, $4$,
$5$, in order.
The power behavior of the form factors at large $t$ can be easily 
understood from perturbative gluon exchange. In order to distribute the 
momentum transfer from the virtual photon
to all three quarks in the nucleon, at least two massless
gluons have to be exchanged. Since each of the gluons has a propagator 
$\sim 1/t$, the form factor has to fall off as $1/t^2$. In the case
of $F_2$, there is additional suppression by $1/t$ since a quark spin 
has to be flipped.

In the form factor analysis, we used spectral functions that lead
exactly to the large-$t$ behavior as given in Eq. (\ref{emff:fasy1}).
The spectral functions of the pertinent form factors are separated 
into a hadronic (meson pole) and a quark (pQCD) component as follows,
\begin{eqnarray}
\label{emff:ffism}
& &F_i^s (t)= \tilde{F}_i^s (t) L(t) \\
& &= \Bigg[ \sum_{V_s}
\frac{a_i^{V_s} \, L^{-1}(m^2_{V_s})}{m^2_{V_s} - t }
\Bigg] \, \Bigg[ \ln \Bigg( \frac{\Lambda^2 - t}{Q_0^2} \Bigg)
\Bigg]^{-\gamma}\,,\nonumber \\
& & F_i^v (t)
= \Bigg[\tilde{F}_i^\rho (t) + \sum_{V_v}
\frac{a_i^{V_v} \, L^{-1}(m^2_{V_v})}{m^2_{V_v} - t }
\Bigg] \, L(t) \,,\nonumber
\end{eqnarray}
where $F_i^\rho(t) = \tilde{F}_i^\rho (t)\, L(t)^{-1}$
parametrizes the two-pion contribution (including
the $\rho$) in terms of the pion form
factor and the P-wave $\pi \pi \bar N N$ partial wave amplitudes
in a parameter-free manner (for details see \cite{MMD96}).
Furthermore, the parameter $\Lambda$ separates the hadronic from the
quark contributions. The fits performed in \cite{MMD96,HMD96} are rather
insensitive to the presence of the logarithmic factor in the spectral
functions. The factor
$L(t)$ in Eq.~(\ref{emff:ffism}) contributes to the spectral functions
for $t > \Lambda^2$ and in some sense parametrizes the intermediate
states in the pQCD regime.
The particular logarithmic form has been chosen for convenience.
More important is the power behavior of the form factors, which
leads to superconvergence relations of the form
\begin{equation}
\int_{t_0}^\infty {\rm Im}\, F_{1,2} (t) dt =0\,,\ldots\,.
\end{equation}
The asymptotic behavior of Eq.~(\ref{emff:fasy1}) is obtained by
choosing the residues of the vector meson pole terms such that the
leading terms in the $1/t$-expansion cancel. 

The number of isoscalar and isovector poles in
Eq.~(\ref{emff:ffism}) is determined
by the stability criterion discussed in detail in \cite{Sab80}.
In short, we take the minimum number of poles necessary to fit the data. 
Specifically, we have three isoscalar and three isovector poles.
Our fit includes all measured form factor data in the space-
and timelike regions. We stress that we are keeping the number
of meson poles fixed in order not to wash out the predictive power.
Due to the various constraints (unitarity, normalizations, superconvergence
relations), we end up with only three free parameters since two (three) of
the isovector (isoscalar) masses can be identified with the masses of
physical particles. We also have performed fits with more poles; these
will not be be mentioned and we refer the reader to
Refs.~\cite{MMD96,HMD96}. 

\subsection{Timelike Data}

We should also comment on the extraction of the timelike form
factor data from experiment. At the nucleon-antinucleon threshold, one has
by definition
\begin{equation}
G_{M} (4 M^2) = G_{E} (4 M^2) \quad ,
\label{emff:Gthr}
\end{equation}
while at large momentum transfer one expects the magnetic form factor to
dominate. The form factors are complex
in the timelike region, since several physical thresholds are open.
Separating $|G_{M}|$ and $| G_{E}|$ unambiguously
from the data requires a measurement of the angular distribution,
which is difficult. 
In most experiments, it has been assumed that either $|G_{M}|= |G_{E}|$ 
or $|G_{E}|= 0$ instead. Most recent data have been presented for the 
magnetic form factors. 

\subsection{Fit Results}

We discuss two of the various fits performed in 
Refs.~\cite{MMD96,HMD96} in detail: fit 1 is a fit to all spacelike data,
while fit 2 also includes the data in the timelike region.
\begin{figure*}[t]
\centering
\includegraphics*[width=118mm]{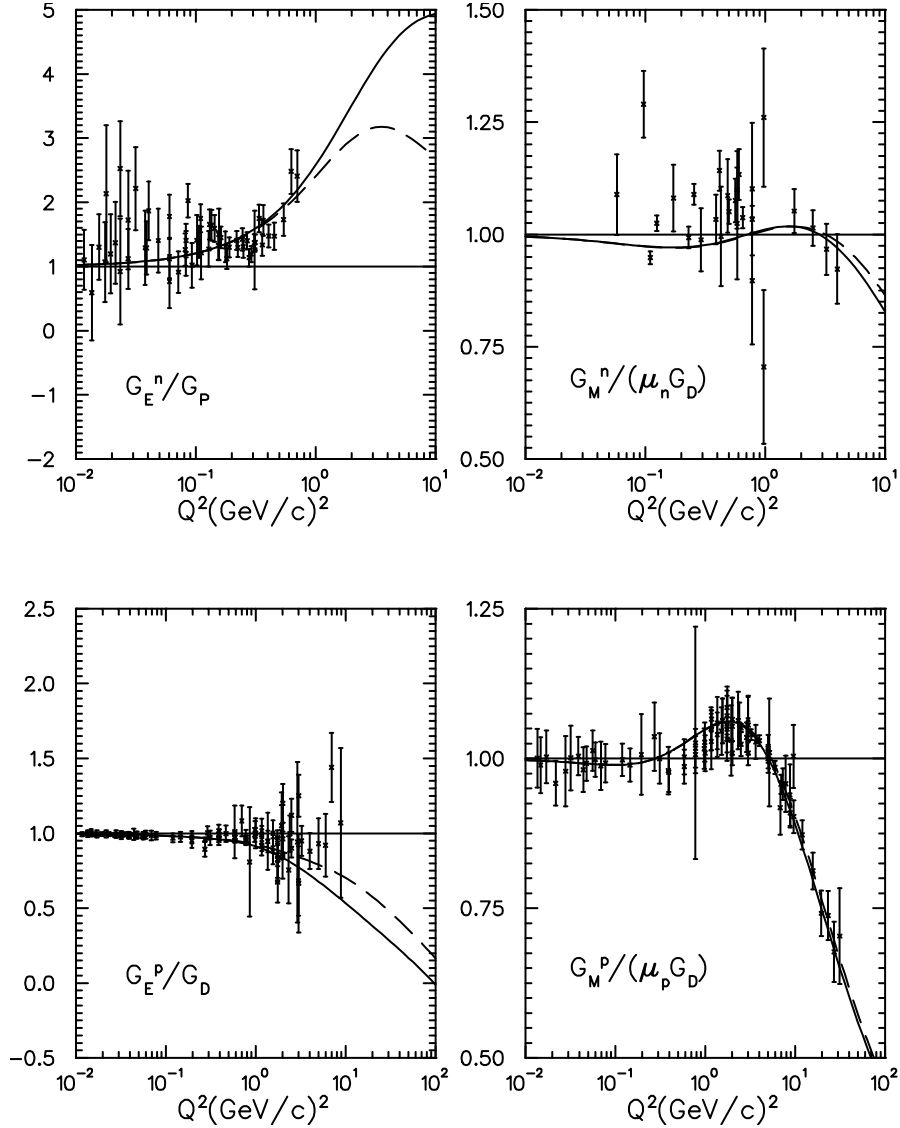}
\caption{Normalized nucleon form factors for spacelike momentum transfer
($Q^2=-t>0$). $G_M^n$, $G_E^p$, and $G_M^p$ are normalized to the dipole fit,
while $G_E^n$ is normalized to the Platchkov fit \protect\cite{Pla90} 
adjusted to give the correct radius. Dashed lines: fit 1, spacelike data only. 
Solid lines: fit 2, including timelike data.} 
\label{raum}
\end{figure*}
We stress that in all fits, the four form factors have been fitted 
simultaneously. This procedure ensures that the results for
different form factors are compatible and allows us to test the
consistency of different data sets.
For fit 1, the vector meson poles in the isoscalar channel are  
the $\omega(782)$, $\phi(1020)$, and $\omega(1600)$. In the 
isovector channel, we have the $\rho(1450)$ and $\rho(1690)$. The
third isovector pole is tightly fixed by the constraints; 
its mass turns out to be close to the $\rho(1690)$.
In fit 2, all masses are the same, except for the isovector pole
at $\sqrt{t}=1690$ MeV which is shifted to $\sqrt{t}=1850$ MeV.
This choice is motivated by the FENICE data \cite{Ant98} which favor
an isovector resonance slightly below the nucleon-antinucleon 
threshold.\footnote{
Note, however, that despite the fact that the vector meson poles 
can be identified with physical vector mesons, it is not clear 
whether this interpretation holds for the higher mass poles. 
If one tries to fit the masses of these poles, they are not
well constrained by the data. In any case, these poles are a
convenient parametrization of the spectral strength in the high
mass region.}
A detailed listing of all fit parameters and the data points
included can be found in Refs.~\cite{MMD96,HMD96}.

In Fig.~\ref{raum}, we compare both fits to the world data in
the spacelike region. 
The dashed lines indicate fit 1, the solid lines indicate fit 2. 
Both fits give a good description of the world data in the spacelike
region, differing only for momentum transfers where the form factors are
not constrained by the data. From fit 1, we extract the following
nucleon radii \cite{MMD96,HMD96},
\begin{equation}
r_E^p = 0.847\mbox{ fm}\,,\; r_M^p = 0.853\mbox{ fm}\,,\;
r_M^n = 0.889\mbox{ fm}\,.
\end{equation}
From the residues of the two lowest isoscalar poles, we can extract
the $\omega NN$ and $\phi NN$ coupling constants \cite{MMD96,HMD96},
\begin{eqnarray}
& &\frac{g_{\omega NN}^2}{4\pi}=34.6\pm 0.8\,, \quad 
\kappa_\omega= -0.16 \pm 0.01 \,, \quad\\
& &\frac{g_{\phi NN}^2}{4\pi}=6.7\pm 0.3\,, \quad 
\kappa_\phi= -0.22 \pm 0.04 \,, \nonumber
\end{eqnarray}
where $\kappa_V$ is the tensor-to-vector coupling ratio.
Both the results for the radii and the vector meson couplings 
are similar to the earlier analysis of Ref.~\cite{Hoe76}.

We now turn to the timelike region.
\begin{figure}[t]
\includegraphics*[width=80mm]{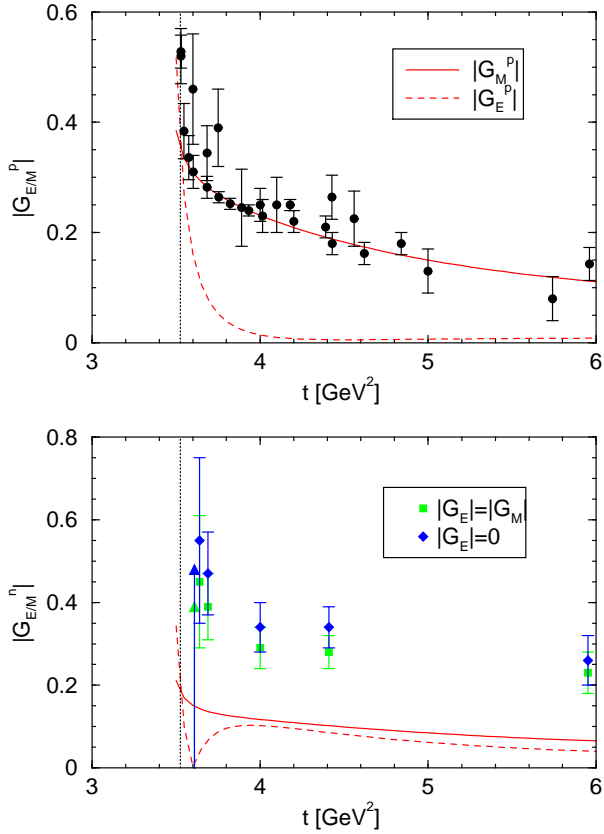}
\caption{Fit 2 to both spacelike and timelike data for timelike
momentum transfer. Vertical dotted line indicates nucleon-antinucleon 
threshold. Solid lines show $|G_M|$, dashed lines show $|G_E|$. 
All neutron data are from the FENICE experiment \protect\cite{Ant98}.} 
\label{zeit}
\end{figure}
In Fig.~\ref{zeit} we show the results of fit 2 compared to the 
world data. Note that all neutron data are from the FENICE
experiment \cite{Ant98}. The solid lines show $|G_M|$,
while dashed lines show $|G_E|$. All proton points are for $|G_M|$ as 
discussed above. The proton magnetic form factor is well described by the
fit. The FENICE data for the neutron magnetic form factor have been 
analyzed under both the assumption $|G_E|=|G_M|$ (squares) and 
$|G_E|=0$ (diamonds). The latter hypothesis is favored by the 
measured angular distributions \cite{Ant98}. Neither data set can be
described by the fit. Even when the experimental
uncertainties of the FENICE data are reduced by a factor of 10
or more, the fit cannot be forced through the data. It is possible that
the spectral function simply has not enough freedom to account for 
the FENICE data and additional poles have to be introduced. It is also
surprising that the neutron form factor is larger in magnitude than the 
proton one, as pQCD predicts asymptotically equal magnitudes.
In any case, there is interesting physics in the timelike neutron form 
factors and more data is called for.
Finally, it is interesting to note that
fit 2 predicts a zero in the neutron electric form factor close to
threshold.

\section{NUCLEON FORM FACTORS IN ChPT}

Chiral perturbation theory (ChPT) is a low-energy effective field theory
of the Standard Model. ChPT focuses on the universal, low-energy
aspects of the physical system. All sensitivity to short-distance
physiscs is captured in a small number of low-energy constants.
ChPT allows for a systematic and model-independent
calculation of low-energy observables by means of an expansion in small 
momenta and quark masses. Particularly important is the spontaneously broken
chiral symmetry of QCD in the limit of vanishing quark masses, which 
guarantees that this expansion is well behaved and severely constrains
the form of the effective Lagrangian.

The first systematic investigation of nucleon form factors in the 
relativistic version of ChPT was given in Ref.~\cite{GSS88}. 
This approach, however, has problems
with the expansion in small momenta since the momenta of the nucleons always
contain the large nucleon mass. If one uses a nonrelativistic formulation
for heavy nucleons, this problem can be circumvented. 
This Heavy Baryon ChPT is generally very successful for
one-nucleon observables. The nucleon form factors were studied 
within HBChPT in Refs.~\cite{BKK92}.  
It was found that the chiral description already 
breaks down at relatively small momentum transfers of $-t=Q^2\sim 0.2$
GeV$^2$. Furthermore, the nonrelativistic expansion does not
reproduce the correct singularity structure of the isovector
spectral functions. This problem can be overcome in the recently
proposed \lq\lq infrared regularization'' of Ref.~\cite{BeL99}, which
is manifestly Lorentz-invariant. Since it is relativistic, this approach 
keeps the correct singularity structure in the low-energy domain
and is expected to improve convergence of the chiral expansion as well.
Recently, this method was applied to the electromagnetic
form factors of the nucleon by Kubis and Mei{\ss}ner (KM) \cite{KuM01}.
In the remainder of this section, we briefly review their
results.

\begin{figure}[t]
\includegraphics*[width=78mm]{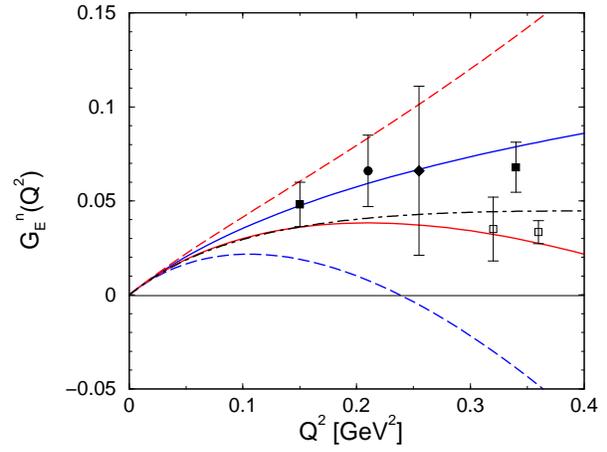}
\caption{$G_E^n$ in relativistic ChPT
(solid lines) to third (upper curve) and fourth order (lower curve).
For comparison: heavy baryon approach (lower/upper dashed line: 
third/fourth order), dispersion analysis (dot--dashed curve),
and some recent experimental points \protect\cite{emdata}.} 
\label{GEn_km}
\end{figure}

Fig.~\ref{GEn_km} shows the electric form factor
of the neutron obtained by KM in relativistic ChPT using 
infrared regularization.
For comparison, the heavy baryon result and the dispersion
analysis discussed above (which can be taken as a representation of
the world data) are also shown. 
The data points are from recent polarization experiments \cite{emdata}.
The relativistic ChPT clearly improves over the heavy baryon calculation,
which already breaks down around $Q^2=0.1$ GeV$^2$. The resummation of 
the $1/M$ terms in the relativistic approach includes recoil
corrections and considerably improves the
convergence of the expansion \cite{KuM01}. 

\begin{figure}[t]
\includegraphics*[width=75mm]{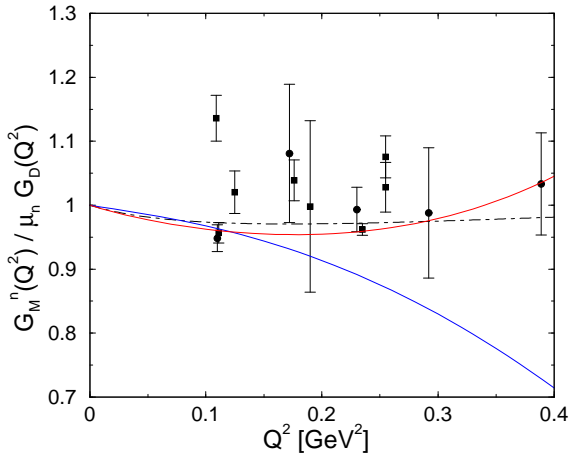}
\caption{$G_M^n$ normalized to the dipole fit
in relativistic ChPT including vector mesons
(solid lines) to third (lower curve) and fourth order (upper curve).
For comparison: world data (cf.\ Refs.~\cite{MMD96,HMD96}) and 
dispersion analysis (dot--dashed curve).}
\label{GMn_dip}
\end{figure}

In the case of the large dipole like form factors $G_M^n$, $G_E^p$,
and $G_M^p$, however, the relativistic method does not give an
improvement over the heavy baryon expansion. These form factors
contain large curvature terms that are not reproduced up to
fourth order in the chiral expansion. 
It is well known that vector mesons contribute
significantly to the nucleon form factors and generate the missing 
curvature terms. In ChPT these contributions are captured in the
low-energy constants. The curvature terms, however, only appear in higher 
orders in the chiral expansion. Since the vector meson couplings to the 
nucleon are known from the dispersion analysis described above 
\cite{MMD96,HMD96}, it is both reasonable and economical to include dynamical 
vector mesons in the theory as no unknown low-energy constants enter.
Following this spirit, KM include the $\rho(770)$, $\omega(782)$, and
$\phi(1020)$ as dynamical fields using the technology of antisymmtric
tensor fields (see Ref.~\cite{KuM01} for details). Using the 
mechanism of resonance saturation, the low-energy constants in the 
theory without vector mesons $L_i$ are split into a vector meson
contribution plus a small remainder $\tilde{L}_i$ which has to be refit, 
\begin{equation}
L_i \to \tilde{L}_i + \sum_V \frac{a_i^V}{m_V^2 -t}\,.
\end{equation}
Keeping the full $t$-dependence in the vector meson contribution
amounts to resumming a selected class of higher order diagrams
in the chiral expansion. 
Fig.~\ref{GMn_dip} shows the results for $G_M^n$
obtained by KM using this method.
Their curves compare well with the world data and the dispersion result.
KM find that the validity of the chiral expansion can be extended up to 
$Q^2\approx 0.4$ GeV$^2$ using this partial resummation.
Similar results hold for the other two dipole-like form factors
$G_M^p$ and $G_E^p$. Furthermore, the good agreement for $G_E^n$
is not destroyed by the inclusion of vector mesons. 

To summarize, Kubis and Mei{\ss}ner have shown that one can
obtain an accurate description of all four nucleon form factors 
up to $Q^2\approx 0.4$ GeV$^2$ within the framework of relativistic 
ChPT and dynamical vector mesons \cite{KuM01}. 
An extension of this work to $SU(3)$ can be found in Ref.~\cite{KuM01b}.

\section{CONCLUSIONS}

Dispersion theory and ChPT are two successful approaches to the em form 
factors of the nucleon that mutually complement each other. Dispersion
theory consistently describes the form factors over the whole
range of momentum transfers in both the spacelike and timelike regions,
while ChPT gives a model independent description at
low spacelike $t$ that explicitly incorporates chiral symmetry. 
The work of KM shows the promise of extending the chiral description to
higher momentum transfers by merging dispersive methods and ChPT \cite{KuM01}.

The neutron form factors in the timelike region show some interesting
features \cite{Ant98}. 
In particular, the magnitude of the neutron form factors close to the
nucleon-antinucleon threshold is larger than expected and at present 
cannot be described by dispersion theory.
More precise data for the em form factors of the nucleon in the timelike
region is called for.

\section{ACKNOWLEDGEMENTS}

I thank R.J.~Furnstahl and U.-G.~Mei{\ss}ner for a careful
reading of the manuscript and B.~Kubis and U.-G.~Mei{\ss}ner for 
providing their figures.
This work was supported under NSF Grant No.\ PHY--9800964.

\end{document}